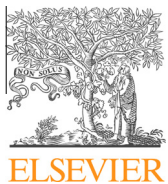

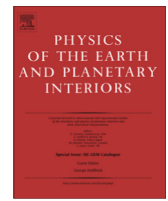

# Core flows and heat transfer induced by inhomogeneous cooling with sub- and supercritical convection

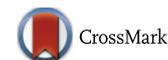

W. Dietrich [a,*], K. Hori [a,b], J. Wicht [c]

[a] Department of Applied Mathematics, University of Leeds, West Yorkshire, United Kingdom
[b] Insitute for Space-Earth Environmental Research, Nagoya University, Nagoya, Japan
[c] Max Planck Institute for Solar System Research, Göttingen, Germany

## ARTICLE INFO

## ABSTRACT

The amount and spatial pattern of heat extracted from cores of terrestrial planets is ultimately controlled by the thermal structure of the lower rocky mantle. Using the most common model to tackle this problem, a rapidly rotating and differentially cooled spherical shell containing an incompressible and viscous liquid is numerically investigated. To gain the physical basics, we consider a simple, equatorial symmetric perturbation of the CMB heat flux shaped as a spherical harmonic $Y_{11}$. The thermodynamic properties of the induced flows mainly depend on the degree of nonlinearity parametrised by a horizontal Rayleigh number $Ra_h = q^*Ra$, where $q^*$ is the relative CMB heat flux anomaly amplitude and $Ra$ is the Rayleigh number which controls radial buoyancy-driven convection. Depending on $Ra_h$ we identify and characterise three distinctive flow regimes through their spatial patterns, heat transport and flow speed scalings: in the linear *conductive* regime the radial inward flow is found to be phase shifted 90° eastwards from the maximal heat flux as predicted by a linear quasi-geostrophic model for rapidly rotating spherical systems. The *advective* regime is characterised by an increased $Ra_h$ where nonlinearities become significant, but is still subcritical to radial convection. There the upwelling is dispersed and the downwelling is compressed by the thermal advection into a spiralling jet-like structure. As $Ra_h$ becomes large enough for the radial convection to set in, the jet remains identifiable on time-average and significantly alters the global heat budget in the *convective* regime. Our results suggest, that the boundary forcing not only introduces a net horizontal heat transport but also suppresses the convection locally to such an extent, that the net Nusselt number is reduced by up to 50%, even though the mean CMB heat flux is conserved. This also implies that a planetary core will remain hotter under a non-homogeneous CMB heat flux and is less well mixed. A broad numerical parameter investigation regarding Rayleigh number and the relative heat flux anomaly further fosters these results.

Crown Copyright © 2016 Published by Elsevier B.V. This is an open access article under the CC BY license (http://creativecommons.org/licenses/by/4.0/).

## 1. Introduction

The cooling of the liquid iron cores of terrestrial planets is due to radial heat transport towards the core mantle boundary (CMB) via heat conduction and in case the entropy gradient is sufficiently negative supported by buoyancy driven convection. The lateral variation of heat conducted out of the core and hence through the CMB $q_{cmb}$ is mainly controlled by the lower mantle temperature pattern $T_{lm}(\theta, \phi)$, such that

$$q_{cmb}(\theta, \phi) = k \frac{T_{lm}(\theta, \phi) - T_{core}}{\delta_{cmb}}, \qquad (1)$$

where $k$ is the thermal conductivity and $\delta_{cmb}$ the thickness of the thermal boundary layer at the bottom of the mantle, respectively. To first order, the core temperature is rather uniform due to a much more efficient conductive and convective heat transport therein. If the heat transport is only via conduction, thermal inhomogeneities at the CMB are thought to drive baroclinic flows (Zhang and Gubbins, 1992), whereas in a convecting core lateral variations of convective vigour, the dynamo process and the stimulation of mean horizontal flows are expected.

The importance of thermal coupling between Earth's mantle and core due to inhomogeneities of the lower mantle temperature was first suggested by Bloxham and Gubbins (1987). Seismological evidence for the non-homogeneous CMB heat flux came from the mantle tomography (e.g. Masters et al., 2000) and the detection of LLSVPs (large low shear velocity provinces) (Yuen et al., 1993), which are associated with lower local mantle temperatures, but it was also reported that only for an isochemical mantle the variation of shear wave velocity and the temperature is linear (Nakagawa and Tackley, 2008). Such a mantle control has been suggested to influence the geomagnetic secular variation (Bloxham, 2000; Davies et al., 2008; Olson et al., 2010; Aubert et al., 2013) and field strength (Takahashi et al., 2008), concentrate magnetic flux patches (Olson and Christensen, 2002; Amit et al., 2010), lock the (usually drifting) core convection (Davies et al., 2009) and dynamo action (Willis et al., 2007) but also introduce mean large scale flows (Gibbons et al., 2007). Furthermore the CMB heat flux anomalies are also reported to affect the buoyancy flux from the inner core (Aubert et al., 2008; Amit and Choblet, 2009; Gubbins et al., 2011). Note, in contrast to most of these dynamo studies, we focus on the hydrodynamic aspects and hence exclude magnetic fields.

Mantle induced CMB heat flux variations are also expected to influence core flows and the magnetic field generation process of other terrestrial planets. For example the rather localised present-day magnetisation on Mars was targeted to be explained by an equatorial antisymmetric mantle-induced CMB heat flux anomaly of variable strength (Stanley et al., 2008; Amit et al., 2011; Dietrich and Wicht, 2013; Monteux et al., 2015). These studies report the induction of a global magnetic field with strong equatorially asymmetric intensity reminiscent of the recently measured distribution of magnetised crust on the surface of Mars (Acuña et al., 1999). Also the equatorial asymmetry of Mercury's magnetic field (Cao et al., 2014; Wicht and Heyner, 2014) and the axisymmetry of Saturn's magnetic field (Stanley, 2010) were investigated in the framework of a non-homogeneous CMB heat flux.

In addition, terrestrial exoplanets orbiting their host star in close proximity and typically in a synchronous orbit will receive strong stellar irradiation at the near side with a latitudinal maximum at the equator. Most likely, the heating–cooling dichotomy at the surface will result in a smooth azimuthal varying, equatorial symmetric thermal forcing pattern at the CMB reminiscent of which we use here ($Y_{11}$). There have been attempts to model mantle convection under such a specific heating pattern (Gelman et al., 2011) suggesting the development of a single-plume mantle convection mode. Hence the core flows and the probable induction of a core dynamo in a tidally locked terrestrial exoplanet might be significantly influenced by the enormous difference of stellar irradiation between the near and far side.

The problem of a differentially cooled and rapidly rotating fluid shell has received much attention during the last decades. Zhang and Gubbins (1992) first investigated the pure effect of laterally varying temperature at the outer boundary in the absence of radial convection. The numerical results revealed that the radial inflows are *not* found where the local outer boundary temperature is lowest hence cooling most efficient, but they are phase shifted by a quarter of the azimuthal wavelength of the thermal anomaly to where the *azimuthal gradient* of the temperature is maximal. This results from the vorticity balance between Coriolis and buoyancy terms frequently referred to as a thermal wind balance, but more specifically is a Sverdrup balance in the geophysical fluid dynamics (Pedlosky, 1979, see also below). This implies, that in a rotation dominated system thermal anomalies induce vorticity rather than flows directly. This holds as long as the flows are assumed to be rotation dominated, inviscid and nonlinearities due to temperature advection or inertia are negligible (Gibbons et al., 2007) and is hence not found in models with small rotation rates or infinite Prandtl number (Sun et al., 1994; Zhang and Gubbins, 1996; Davies et al., 2009). A mathematically more straight-forward analysis of the linear quasi-geostrophic model (Busse annulus) by Yoshida and Hamano (1993) including the effect of the magnetic fields, confirmed the azimuthal phase shift when there is no magnetic field altering the leading order force balance.

For a physically more realistic model of the Earth's core, experiments by Sumita and Olson (1999) perturbed the outer boundary of a vigorously convecting and rapidly rotating spherical shell with a local anomalous heat flux. There was also a phase shift between local minimum temperature and position of the inward flow reported, however due to the strongly nonlinear driving it was deformed in a jet-like structure spiralling inwards. As the characteristic hydrodynamic numbers (e.g. Ekman or Reynolds number) in experiments are closer to real planets than those accessible by numerical models, this might reflect better the situation in a realistic planetary core. In a follow-up study (Sumita and Olson, 2002) a broad parameter survey was reported, featuring a detailed description of the flow and scaling relations of how the heat flux anomaly, pattern and strength affects the induced flow amplitudes in azimuthal and radial direction. Typically these experiments are set in a strongly convective regime, where a rather localised and very strong heat flux anomaly induces a sharp front separating the cold east from the hot west and strong azimuthal flows connecting them.

As some of the reported effects are due to complex interactions between core convection, boundary forcing and magnetic fields, a clear physical interpretation might not be always possible. Thus to gather clearer insights we limit this study to the hydrodynamic aspects for the sake of a more coherent physical description. Starting from an analytical formulation of the linear theory, we numerically model core flows induced by thermal CMB inhomogeneities for cores subcritical to buoyancy driven core convection (baroclinic flows) and compare them to models featuring radial convection. Note, also the baroclinic flows obey strong nonlinear features, such as bending or compression of the emerging inward jet. As the various linear and nonlinear flow regimes have been studied only individually, our main focus is to distinguish them and discuss their properties.

More precisely we modify the outer boundary heat flux of rapidly rotating spherical shell with a smoothly varying pattern along azimuthal and latitudinal direction. For simplicity the heating of the shell is exclusively supplied by a constant internal heat source modelling secular cooling of the shell. It was shown that internal heated systems are most sensitive to inhomogeneities at the outer boundary (Hori et al., 2014), as the strongest temperature gradients are typically found close the outer boundary. The specific shape is an anomaly of the outer boundary heat flux of spherical harmonic degree and order one ($Y_{11}$). This purely equatorially symmetric and non-axisymmetric pattern was taken mainly for application to terrestrial exoplanets orbiting their host star in synchronous rotation. Note, today's Earth CMB heat flux variation is dominated by a sectorial spectral mode $Y_{22}$, but the core convection is mainly driven from compositional buoyancy release at the inner core boundary. Hence if applied to real planetary systems, our models best describe cores of terrestrial planets before inner core nucleation. Apart from the different azimuthal length scales, our linear results might be applicable to a general sectorial anomaly pattern as shown below on Section 3.1. However, for the nonlinear results the interaction between several sectorial heat flux anomalies are beyond this study. Emphasis is put on a clear physical description of the induced flow structures and the radial (and horizontal) heat transport.

We also vary the amplitude of the CMB heat flux variation $q^*$ relative to the mean value as it controls the strength of the boundary forcing to first order and is not even well known for the Earth's core. We are interested in how sensitive convection is to a thermal boundary anomaly of variable strength. Studying $q^*$ might provide a smooth transition between models with homogeneous and heterogeneous CMB heat flux. Note, we limit $q^*$ to unity to avoid any heat flux into the core which might not be consistent with our model assumptions. A fundamental property of the here applied Boussinesq-approximation is that all temperatures and heat fluxes are understood as superadiabatic fluctuations on top of the (excluded) adiabatic background state of heat conduction. Hence a $(q^* > 1)$-heat anomaly describes locally a superadiabatic heat flux *into* the core in a Boussinesq-model, whereas the physical meaning is a subadiabatic heat flux *out* of the core.

In a stably stratified, rapidly rotating fluid shell motions are introduced due to baroclinic torques emerging from the misalignment of surfaces of constant pressure and constant density. As we consider a fluid in Boussinesq approximation, the equation of state simplifies to a linear relation between density and temperature due to thermal expansion. Hence the leading order vorticity balance is given by the curl of the Navier–Stokes equation

$$-2\Omega \frac{\partial \vec{v}}{\partial z} = \rho \alpha \nabla \times (T\vec{g}), \tag{2}$$

where $\Omega$ is the rotation vector, $\vec{v}$ the (dimensional) flow velocity, $\rho$ the reference density, $\alpha$ the thermal expansivity, $T$ the temperature and $\vec{g}$ the linear increasing gravity. Physically this implies, that any horizontal temperature variation introduces a finite flow (baroclinicity). However, if the temperature varies only along radius, pressure and density are aligned and the flow comes to a rest (hydrostatic equilibrium).

The modelling strategy of this paper attempts to cover all possible linear and nonlinear flow states. We start off with deriving the analytical description like Yoshida and Hamano (1993), but for spherical shells rather than a cylindrical annulus, adding the ageostrophic part of the flow and without magnetic fields. These results are verified by numerical simulations set in the almost linear (conductive) regime (Section 3.1). Then we increase the driving to reach a nonlinear regime, that is yet not supercritical to radial convection and discuss the properties of the emerging inward jet (Section 3.2). To understand how radial convection and the inward jet interact, we finally analyse the convective regime where radial convection is strongly participating. Furthermore it is an open question, how and to which extent boundary forcing alters the global heat transfer budget. Therefore an investigation of the three-dimensional redistribution of heat and the mean heat transport properties such as the Nusselt number is described in Section 4. We also parametrise the influence of the vigour of convection and the various relative forcing amplitudes.

## 2. Methods and model

The outer core spherical shell with outer and inner radius, $r_o$ and $r_i$, is assumed to be filled with an incompressible and viscous liquid. Within the Boussinesq-approximation the evolution of the fluid flow is governed by the dimensionless Navier–Stokes equation:

$$E\left(\frac{\partial \vec{u}}{\partial t} + \vec{u} \cdot \vec{\nabla}\vec{u}\right) = -\vec{\nabla}\Pi + E\nabla^2\vec{u} - 2\hat{e}_z \times \vec{u} + \frac{RaE}{r_o Pr}\hat{e}_r T, \tag{3}$$

where $\vec{u}$ is the incompressible velocity field ($\vec{\nabla} \cdot \vec{u} = 0$), $\Pi$ the non-hydrostatic pressure, $\hat{e}_z$ the direction of the rotation axis and $T$ the super-adiabatic temperature fluctuation on top of a hydrostatic equilibrium state. We use the shell thickness $D = r_o - r_i$ as length and the viscous diffusion time $D^2/\nu$ as time scale. The mean super-adiabatic CMB heat flux density $q_o$ serves to define the temperature scale $q_o D/c_p \rho \kappa$. Further, $\nu$ is the viscous diffusivity, $\rho$ the constant background density, $\Omega$ the rotation rate, $\kappa$ the thermal diffusivity and $c_p$ the heat capacity.

The non-dimensional control parameters are the Prandtl number $Pr = \nu/\kappa$ as the ratio between the viscous and thermal diffusivities, the Ekman number $E = \nu/\Omega D^2$ relates viscous and rotational time scale and the Rayleigh number $Ra = \alpha g q_o D^4/c_p \rho \nu \kappa^2$ controls the vigour of convection.

The evolution of the thermal energy is affected by thermal diffusion and advection along with the flow, such that

$$\frac{\partial T}{\partial t} + \vec{u} \cdot \vec{\nabla}T = \frac{1}{Pr}\nabla^2 T + \epsilon, \tag{4}$$

where $\epsilon$ is a uniform heat source density.

We assume non-penetrative and no-slip mechanical boundary conditions because the mantle and inner core are solid. When core convection is driven by secular cooling only, we set the heat flux at the outer boundary to unity ($q_o = 1$) and apply zero flux on the inner core boundary ($q_i = 0$). From the modelling perspective, the system is then powered exclusively by the internal heat source $\epsilon$ and a simple balance between $\epsilon$ and the mean CMB heat flux $q_o$ is given by

$$\epsilon = \frac{1-\beta}{1-\beta^3} \frac{3q_o}{Pr}, \tag{5}$$

where $\beta = r_i/r_o$ is the radius ratio of the spherical shell. Such an internal heating setup models a shell undergoing secular cooling, but neglects compositional and thermal buoyancy arising from inner core solidification.

The thermal effect of inhomogeneous cooling is modelled with a spherical harmonic of degree and order unity ($Y_{11}$) with variable amplitude $q^* = [0 \dots 1.0]$, such that

$$q_{cmb}(\theta, \phi) = q_o(1 + q^* Y_{11}). \tag{6}$$

The boundary anomaly proportional to $Y_{11}$ provides a smaller than average heat flux in the sectors I and IV (see Fig. 1). Actually, for $q^* = 1$ it is exactly zero at $\phi = 0$. In the sectors II and III the heat flux is enhanced, with a maximum of $2q_o$ at $\phi = \pi$. In between the spherical harmonic description guarantees a smooth variation (see also Fig. 1b).

The emerging horizontal difference in convective intensity is measured in terms of a horizontal Rayleigh number $Ra_h$ (Willis et al., 2007; Monteux et al., 2015), such that

$$Ra_h = q^* Ra. \tag{7}$$

As we use the Boussinesq-approximation, the horizontal Rayleigh number shall not exceed the (radial) reference Rayleigh number of the homogeneous model. This is assured when keeping $q^* \leqslant 1$.

The combination of radial and horizontal temperature gradients will lead to a complex superposition of various flows. Fig. 1 illustrates the equatorial plane of the three-dimensional model and clarifies the nomenclature used throughout the paper. $\Omega$ gives the sense of rotation, the azimuthal angle $\phi$ is counted in prograde direction starting from the right hand side. Hence the upper half ($0 \leqslant \phi \leqslant \pi$) is termed *eastern*, the lower *western* hemisphere. For convenience the plane is further subdivided into the four azimuthal sectors (I to IV). The heat flux anomaly is positioned such that, the hottest mantle feature is on the right at $\phi = 0$, where then also the smallest amount of heat is extracted from the core as indicated by the small red arrow in the figure. Thus the largest CMB heat flux is at the opposite point (at $\phi = \pi$). We expect that in the absence of radial convection, the boundary anomaly should drive a mean flow consisting of two cells, a (eastern) cold cyclone and a western hot anticyclone. The cells converge into an upwelling (downwelling) where the CMB heat flux is small (large). However, the dominant Coriolis force will turn the entire flow system. Indeed, as we shall show later, for models featuring radial convection the downwelling is typically shifted by an angle $0° \leqslant \Phi \leqslant 90°$ somewhere in sector III. For clarity, we measure the phase shift $\Phi$ in degrees and generic azimuthal angle $\phi$ in radians. Earlier weakly nonlinear numerical (Zhang and Gubbins, 1992) and linear analytical studies (Yoshida and Hamano, 1993) suggested a phase shift of $\Phi = 90°$ between the heat flux maximum and the radial inflow, whereas for a non-rotating model $\Phi = 0$ would be expected.

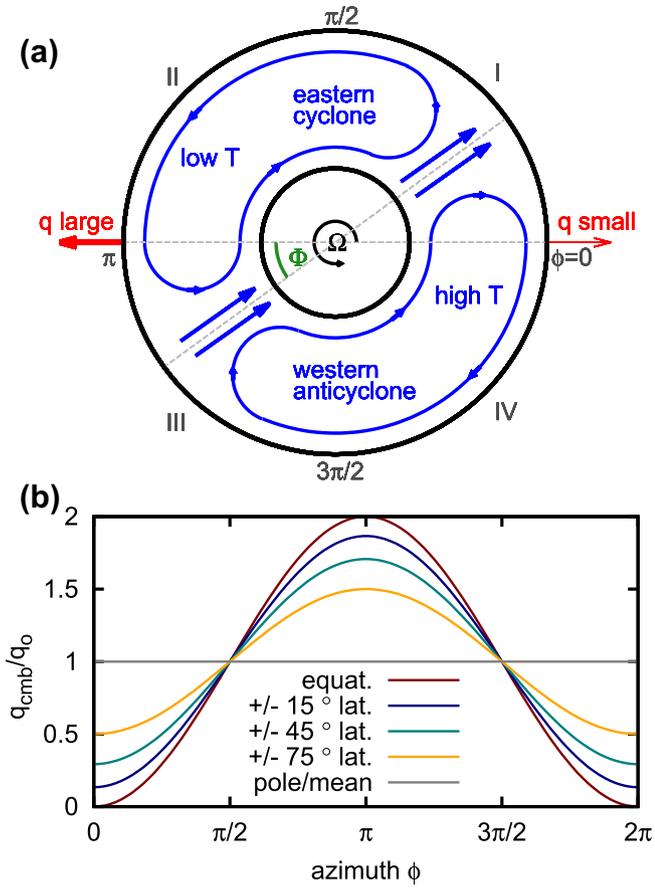

**Fig. 1.** Basic setup and nomenclature (a) and CMB heat flux for various latitudes (b, using $q^* = 1.0$). The horizontally varying outer boundary heat flux is minimal (maximal) at $\phi = 0$ ($\phi = \pi$). In a) $\Phi$ (green) denotes the phase shift of inward flow measured from the $\phi = \pi$. (For interpretation of the references to colour in this figure legend, the reader is referred to the web version of this article.)

We numerically solve the governing equations (Eqs. (3) and (4)) with the pseudo-spectral MagIC3 code (Wicht, 2002). The radial and horizontal resolution is mainly 49 and 288, respectively, but is increased up to 61 and 320 for the computational most demanding cases with high Rayleigh numbers. The radial resolution is not equidistant featuring higher grid point density close to the boundaries. E.g., the thickness of the Ekman layer in the vicinity of the boundaries is given by $\delta/D \approx E^{1/2} = 0.01$ and is resolved with several radial grid points. In addition the spectral decay in the flow spectrum from the largest to the smallest length scales was checked to consistently span three orders of magnitude. In total 80 numerical models with different setups are solved, where each individual case is time integrated from an initial condition taken from a convecting or randomised solution until a steady or statistically stationary state is reached and time-averaged over a viscous diffusion time. A steady state is characterised by an exactly vanishing time derivative, whereas a statistically stationary state is reached when the kinetic energy fluctuates around a constant mean value. For investigating qualitatively the various linear and nonlinear flow regimes, 28 different models covering 8 orders of magnitude in Rayleigh number are calculated, where all other parameters are kept fixed ($q^* = 1.0, E = 10^{-4}, Pr = 1, \beta = 0.35$) if not stated otherwise. For reference we ran 26 equivalent homogeneous models ($q^* = 0.0$). For the combination of Ekman number, Prandtl number, aspect ratio and the internal heating mode with flux boundaries, an equivalent homogeneously cooled model would require a Rayleigh number of at least $Ra_c = 1.6 \cdot 10^6$ for the onset of convection. Note, this value was determined numerically by monitoring the growth or decay of random perturbations from the basic state. For analysing the transition from homogeneous to inhomogeneous boundary heat flux 8 different amplitudes ($q^*$) are tested as well. To get an idea of the Ekman number dependence there are another 16 models with $E = 5 \cdot 10^{-5}$.

## 3. Results

In contrast to the flows introduced by the boundary anomaly, buoyancy driven convective instabilities are opposed by the stabilizing pressure gradient and require a driving stronger than a critical value ($Ra > Ra_c$). According to the vorticity equation (Eq. (2)) there is always a finite flow for inhomogeneous outer boundary heat flux independent of the presence of radial convection. The horizontal Rayleigh number $Ra_h$ serves here as the main parameter to find different regimes. Verifying our analytical results of the linear, spherical problem we start the numerical survey at $Ra_h = 10$, which is strongly subcritical ($Ra_h/Ra_c = 6.25 \cdot 10^{-6}$). As long as the resulting flow amplitudes in terms of the Reynolds or Péclet number, $Re$ or $Pe$, are less than unity the nonlinear temperature advection and inertia terms are negligible, hence this is the linear (conductive) regime. For the nondimensional viscous time scale chosen here, $Re$ is equivalent to the rms flow speed and $Pe = RePr$. Note, as the Prandtl number is fixed throughout the paper to $Pr = 1.0$, Reynolds and Péclet number give identical measures of the flow amplitudes. Increasing $Ra_h$ will increase $Re$ as well, and the thermal advection is found to strongly modify the flow. Hence the nonlinear (advective) regime is defined by nonlinear flows, which are still subcritical to convection. We set the regime boundary to the nonlinear convective regime, at the critical Rayleigh number $Ra_c$ for the homogeneous system ($Ra_h/Ra_c = 1$). Hence a suite of numerical simulations starting with $Ra_h = 10$ and reaching up to $Ra_h = 6.4 \cdot 10^8$ ($Ra_h/Ra_c = 400$) is expected to cover all possible flow states and transitions. As the Ekman number is fixed at $E = 10^{-4}$, a further enhancement of $Ra$ or $Ra_h$ might lead to an inertia-dominated regime, which is unrealistic for planetary applications.

### 3.1. Linear conductive regime

Zhang and Gubbins (1992) predicted for the rapidly rotating, non-magnetic and inviscid limit ($E \rightarrow 0$), that radial flows are determined by the azimuthal temperature gradients rather than the outer boundary radial heat flux. Hence up- and downwellings are not expected to be found where the temperature is largest and smallest, but where the azimuthal temperature gradient is maximised. A mathematical more straight-forward analytical study by Yoshida and Hamano (1993) including magnetic fields confirmed this result within the linear theory of the Busse annulus (Busse, 1970). To understand the physical origin of these flows, we follow the approach of Yoshida and Hamano (1993) but generalise it to spherical shells, add the ageostrophic flows while neglecting the effect of the magnetic field. Numerical simulations with very small $Ra_h$ are used to test the analytical results.

For rapidly rotating systems the thermodynamic balance between Coriolis and buoyancy force can be found by taking the curl of the Navier–Stokes equation (Eq. (3)). For the inertia-less, inviscid and steady limit, the leading order vorticity balance (Eq. (2)) is then given in dimensionless form by

$$-2\frac{\partial \vec{u}}{\partial z} = \frac{RaE}{r_o Pr}\nabla \times (\hat{e}_r T).$$ (8)

We separate the flow into its geostrophic ($\vec{u}^g$) and ageostrophic part ($\vec{u}^a$). As geostrophic flows are by definition invariant along the axis of rotation ($\hat{e}_z$), it is convenient to separate $\vec{u}^g$ and $\vec{u}^a$ using cylindrical coordinates ($s, z, \phi$):

$$\vec{u} = \vec{u}^g(s, \phi) + \vec{u}^a(s, \phi, z).$$ (9)

The geostrophic part is represented by the $z$-averaged flow ($\vec{u}^g = \langle\vec{u}\rangle_z$), where the averaging operator is defined by

$$\langle f\rangle_z(s, \phi) = \frac{1}{2H}\int_{H_-}^{H_+} f(s, \phi, z)dz,$$ (10)

for a generic function $f$ and $H = \sqrt{r_o^2 - s^2}$ is the $s$-dependent height of a spherical shell measured from the equator outside the tangent cylinder (TC). The geostrophic flow can be found by investigating the $z$-averaged $z$-component of Eq. (8):

$$-\left\langle\frac{\partial u_z}{\partial z}\right\rangle_z = \frac{RaE}{2r_o Pr}\langle\hat{e}_z \cdot (\nabla \times \hat{e}_r T)\rangle_z.$$ (11)

The left hand side of the equation can be directly integrated. Incorporating the sloping boundaries conditions (e.g. Busse, 1970; Gillet and Jones, 2006; Hori et al., 2015) gives

$$\int_{H_-}^{H_+}\frac{\partial u_z}{\partial z}dz = u_z(H_+) - u_z(H_-) = -2u_s^g\frac{s}{H}.$$ (12)

For the right hand side, the balance (Eq. (11)) has in general a latitudinal and an azimuthal component, where the latter is antisymmetric with respect to the equator for any equatorial symmetric boundary anomaly and averages out ($\langle\frac{\partial T}{\partial\theta}\rangle_z = 0$). Hence $z$-averaging the RHS of Eq. (11) yields

$$\frac{1}{2H}\int_{H_-}^{H_+}\hat{e}_z \cdot (\nabla \times \hat{e}_r T)dz = \frac{1}{s}\left\langle\frac{\partial T}{\partial\phi}\hat{e}_\theta \cdot \hat{e}_z\right\rangle_z.$$ (13)

Finally evaluating the equation in the equatorial plane, where $s = r$, $u_s^g = u_r^g, \hat{e}_\theta \cdot \hat{e}_z = -1$ gives the solution for the radial flow outside TC

$$u_{s,r}^g = -\frac{r_o^2 - s^2}{2r_o s^2}\frac{RaE}{Pr}\left\langle\frac{\partial T}{\partial\phi}\right\rangle_z.$$ (14)

From this equation it can be seen that the radial geostrophic flow is amplified where the azimuthal gradients of the temperature are large, i.e. a phase shift between the two. Note, this relation for the baroclinic driving of geostrophic radial flows holds for any azimuthal variation of the temperature. This is equivalent to the Sverdrup relation, which is commonly known in oceanography (Pedlosky, 1979). There usually the thermal anomaly is replaced by the curl of the wind stress exerted onto the free surface of an ocean.

The geostrophic azimuthal flow ($u_\phi^g$) will connect the up- and downwellings and is found from the $z$-averaged incompressibility condition in cylindrical coordinates. The solenoidal condition for the geostrophic flow simplifies to:

$$\frac{\partial(su_s^g)}{\partial s} + \frac{\partial u_\phi^g}{\partial\phi} = 0.$$ (15)

Furthermore, by introducing a stream function $\Psi(s, \phi)$ radial and azimuthal flows can be immediately calculated as

$$u_s = \frac{1}{s}\frac{\partial\Psi}{\partial\phi} \quad\text{and}\quad u_\phi = -\frac{\partial\Psi}{\partial s}.$$ (16a, b)

From Eq. (14) we finally obtain

$$\Psi = -\frac{r_o^2 - s^2}{2r_o s}\frac{RaE}{Pr}\langle T\rangle_z.$$ (17)

Hence the azimuthal, geostrophic flow $u_\phi^g$ is not phase shifted relative to, but obeys a more complex radial structure than the temperature field. As can be seen, azimuthal flows converge (diverge) where radial flows are increasing (decreasing) along radius. We therefore expect a prograde (retrograde) azimuthal flow in sectors I and III (sectors II and IV).

We discuss the temperature and the ageostrophic part of the flow as well. As the heat flux anomaly is given in spherical harmonics, switching back to spherical coordinates ($r, \theta, \phi$) is physically and mathematically more suitable, however on the expense of a mild inconsistency with the cylindrical treatment of the geostrophic flow. If temperature advection is negligible, Eq. (4) simplifies to a diffusion equation subject to the outer boundary heat flux variation, which shall be shaped like a general sectorial spherical harmonic $Y_{mm} = \sin(m\phi)P_m^m(\cos\theta)$ with amplitude unity:

$$q_0(\theta, \phi) = 1 + \sin(m\phi)P_m^m(\cos\theta).$$ (18)

As there is no temperature advection considered, the boundary anomaly will diffuse as a temperature anomaly through the entire shell, hence

$$T(r, \theta, \phi) = R(r)\sin(m\phi)P_m^m(\cos\theta),$$ (19)

where $R(r)$ is a undetermined function describing the radial dependence only. The azimuthal and latitudinal gradients can be given as:

$$\frac{\partial T}{\partial\phi} = -R(r)\,m\cos(m\phi)P_m^m(\cos\theta)$$ (20)

$$\frac{\partial T}{\partial\theta} = R(r)\,m\sin(m\phi)\cot\theta P_m^m(\cos\theta).$$ (21)

As a consequence only non-axisymmetric ($m \geq 1$) disturbances of the CMB heat flux drive radial geostrophic flows, where then the azimuthal displacement between the temperature and radial flow is a quarter of the azimuthal variational wave length. Further we note, the $z$-average of Eq. (21) disappears always for sectorial patterns as the integrand ($\cot\theta\,P_m^m(\cos\theta)$) is antisymmetric with respect to the equator.

By definition geostrophic flows are invariant along $z$, hence the ageostrophic flow is found in the spherical components of the Eq. (8)

$$\frac{\partial u_r^a}{\partial z} = 0$$ (22)

$$\frac{\partial u_\theta^a}{\partial z} = -\frac{RaE}{2r_o Pr}\frac{1}{\sin\theta}\frac{\partial T}{\partial\phi}$$ (23)

$$\frac{\partial u_\phi^a}{\partial z} = \frac{RaE}{2r_o Pr}\frac{\partial T}{\partial\theta}.$$ (24)

Hence the ageostrophic part of the flow is found by the temperature gradient along colatitude for the axial variation of the azimuthal flow and along azimuthal angle for the latitudinal flow variation. The azimuthal component (Eq. (24)) is more commonly known as the thermal wind balance in thick spherical shells (Jones, 2007). The ageostrophic flows are entirely determined by Eqs. (22)–(24) hence act only along azimuth $\phi$ and latitude $\theta$, but are non-radial. As the heat flux anomaly gets weaker towards the poles, the latitudinal temperature gradient is either negative where the equator is hotter than the pole ($\phi = 0$) or positive where the equator is colder than the pole (at $\phi = \pi$). A schematic flow is shown in Fig. 2, panel b. That implies (Eq. (24)) a prograde (retrograde) azimuthal flow at the equator in sectors II and III (IV and I), hence the far side (near side) of the planet. The ageostrophic azimuthal flow $u_\phi^a$ is thus consistent with its geostrophic counterpart $u_\phi^g$ and retrograde (prograde) at the front (far) side of the shell in the equatorial plane.

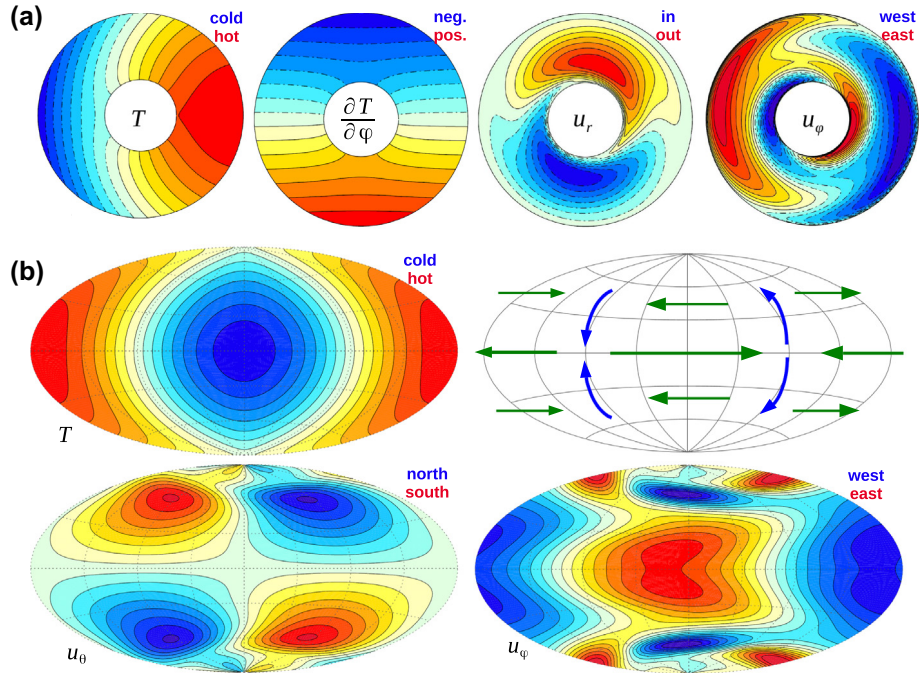

**Fig. 2.** Temperature, geostrophic and ageostrophic flow for the conductive regime ($Ra_h = 10$), where $T$ at the outer boundary equals $-q_{cmb}$. (a) temperature, azimuthal temperature gradient, radial and azimuthal flow (from left to right) in the equatorial plane. (b) spherical surface projections centered at $\phi = \pi$ of the temperature ($T$), the schematic flow expected from the Eqs. (23) and (24) with $u_\theta$ in blue and $u_\phi$ in green, the last row shows simulation results of latitudinal ($u_\theta$) and azimuthal flow ($u_\phi$). Parameters: $E = 10^{-4}, q^* = 1.0, Ra = Ra_h = 10$. (For interpretation of the references to color in this figure legend, the reader is referred to the web version of this article.)

Furthermore, as the temperature decreases (increases) in the eastern (western) hemisphere, the latitudinal flows are expected to converge (diverge) at the equator. Note both flow parts, geostrophic $u_\phi^g(r, \phi), u_\phi^g(r, \phi)$ and ageostrophic $u_\theta^a(\theta, \phi), u_\phi^a(\theta, \phi)$, provide a unique two dimensional closed circulation seeking to equilibrate the heat flux anomaly.

Fig. 2 schematically depicts the expected flow structures and compares to the simulations. Subpanel (a) shows that the numerical results are consistent with the theoretical treatment for the geostrophic flow, where (b) shows the temperature distribution and the horizontal flow components at a spherical surface for the ageostrophic flow. The reversed $u_\theta$ at higher latitudes is the ageostrophic equivalent of the geostrophic return flows closer to the inner core. Whereas in the geostrophic flow an azimuthal temperature gradient drives a $z$-independent radial flow, for the ageostrophic flow this results in an equatorially antisymmetric convergence or divergence of latitudinal flow. Hence equatorward flows are expected in the eastern hemisphere, whereas polewards flows are driven in the western hemisphere. Note, that the ageostrophic flow does not require any radial flows in order to equilibrate the heating anomaly, however the geostrophic part cannot be neglected as it exceeds the ageostrophic part in amplitude and will naturally emerge in any rapidly rotating, boundary forced system. In addition from the Eqs. (14) and (23) it can be seen, that the (purely geostrophic) radial flow ($u_r = u_r^g$) and the (purely ageostrophic) latitudinal flow gradient ($\partial u_\theta / \partial z = \partial u_\theta^a / \partial z$) are both proportional to $\partial T / \partial \phi$ and hence are governed by the azimuthal temperature gradients rather than the temperature itself. For the azimuthal flow ($u_\phi$), the geostrophic part found from the Eq. (15) and the ageostrophic flow gradients defined by Eq. (24) are proportional to the temperature rather than the temperature gradient.

In addition we numerically isolate the geostrophic ($z$-independent) part of the flow resulting from the numerical models by transforming from spherical to cylindrical coordinate representation and measuring the rms flow amplitude by

$$Re_i^g = \sqrt{\frac{1}{V} \int_V (u_i^g)^2 dV} \tag{25}$$

$$Re_i^a = \sqrt{\frac{1}{V} \int_V (u_i - u_i^g)^2 dV}, \tag{26}$$

where the index $i$ stands for either radial ($r$), latitudinal ($\theta$) or azimuthal flow ($\phi$). Here we use the time-averaged flow assuming only the boundary induced flow structures are time-persistent. The geostrophy as the ratio $Re_i^g / Re_i^a$ is calculated from the simulations and listed in the Table 1 for several $Ra_h$. For the latitudinal direction, indeed the geostrophic contribution is much smaller than the ageostrophic as expected from the analytical results. Interestingly this is less clear for the (as expected purely geostrophic) radial flow, where the ageostrophic part is probably larger than expected due to boundary or viscous effects. As evident from Table 1 the mean azimuthal Reynolds number of the geostrophic flow is seven times larger than the ageostrophic.

**Table 1**
Geostrophy of the radial ($Re_r$), latitudinal ($Re_\theta$) and azimuthal ($Re_\phi$) calculated as the ratio of the geostrophic and ageostrophic Reynolds number ($Re_i^g$ and $Re_i^a$, respectively) for the six study cases from Fig. 4. Note, all values are calculated from time averaged or steady flows.

| Regime | $Ra_h$ | $Re_r^g / Re_r^a$ | $Re_\theta^g / Re_\theta^a$ | $Re_\phi^g / Re_\phi^a$ |
|---|---|---|---|---|
| Conductive | 10 | 1.81 | 0.113 | 6.96 |
| Advective | $3 \cdot 10^4$ | 1.81 | 0.108 | 6.89 |
| | $3 \cdot 10^5$ | 1.90 | 0.076 | 5.71 |
| Convective | $5 \cdot 10^6$ | 1.97 | 0.160 | 2.49 |
| | $4 \cdot 10^7$ | 1.73 | 0.212 | 2.12 |
| | $3.2 \cdot 10^8$ | 1.69 | 0.326 | 3.58 |

To visualise the ratio of geostrophic and ageostrophic flow contributions the bending of lines of constant $u_\phi$ is plotted. Fig. 3 depicts the azimuthal flows $u_\phi$, temperature $T$ and the left- and right-hand side of the thermal wind (Eq. (24)) in meridional cuts along $\phi = 0$ and $\phi = \pi$. The rather weak bending confirms the geostrophic flow (at least the azimuthal) exceeds the ageostrophic flow by far. However it can be seen, the weak variation of $u_\phi$ along $z$ is given by the azimuthal thermal wind. Interestingly, an additional shear flow feature not related to the thermal wind appears at the TC. This was suggested to be caused by viscous effects on the geometric discontinuity across the TC creating azimuthal shear flows (Livermore and Hollerbach, 2012).

### 3.2. Nonlinear regimes: advective and convective

As convincing and clear the analytical results are, the simple linear treatment oversimplifies a more realistic scenario. The nonlinear advection terms in the temperature and momentum equation will significantly contribute and complicate flow and temperature once their nondimensional measure (Reynolds and Péclet number) reaches unity. We therefore numerically investigate the effect of an increasing (horizontal) Rayleigh number and characterise the different regimes.

As an overview, Fig. 4 shows the time-averaged flow and temperature distribution found in the equatorial plane for several $Ra_h$ ranging from $Ra_h = 10$ to $3.2 \cdot 10^8$. As extensively discussed above, for the smallest $Ra_h$ (top row) the flow structure is in agreement with numerical and analytical results (Zhang and Gubbins, 1992; Yoshida and Hamano, 1993). Note, a few plots from Fig. 2 are repeated to complete the overview. As introduced in the Section 3.1 an equatorial stream function yields a clearer characterisation. Radial and azimuthal flows can then be deduced by

$$u_r = \frac{1}{r}\frac{\partial \Psi}{\partial \phi} \quad \text{and} \quad u_\phi = -\frac{\partial \Psi}{\partial r}, \tag{27a, b}$$

in spherical coordinates (cf. Eq. (16a,b)). The anticyclonic structure further east of the jet appears negative (red), whereas the cyclonic structure is positive (blue).

To better quantify the different phase relations, azimuthal profiles of mean temperature $T$, azimuthal temperature gradient $\partial T/\partial \phi$ and radial flows $u_r$ are defined by averages with respect to time, radius and colatitude:

$$T(\phi) = \int_{r_i}^{r_o} \int_0^\pi \langle T \rangle_t r^2 \sin\theta \, dr \, d\theta \tag{28}$$

$$dT/d\phi = \int_{r_i}^{r_o} \int_0^\pi \frac{\partial \langle T \rangle_t}{\partial \phi} r \, dr \, d\theta \tag{29}$$

$$u_r(\phi) = \int_{r_i}^{r_o} \int_0^\pi \langle u_r \rangle_t r^2 \sin\theta \, dr \, d\theta \tag{30}$$

$$|u_r'|(\phi) = \int_{r_i}^{r_o} \int_0^\pi \sqrt{\langle u_r^2 \rangle_t} r^2 \sin\theta \, dr \, d\theta, \tag{31}$$

where $\langle f(r, \theta, \phi) \rangle_t = 1/\Delta t \int f(\theta, \phi, t) dt$ is the time average operator. Note, $|u_r'|(\phi)$ is derived as the time averaged intensity of non-axisymmetric radial flow in individual snapshots, whereas $u_r(\phi)$ is the remaining radial flow after time-averaging.

Fig. 5 plots the azimuthal profiles of $T, dT/d\phi, u_r$ and $|u_r'|$. The black curves correspond to the upper case ($Ra_h = 10$, conducting regime) from Fig. 4 and show nicely the anti-correlation between the radial flow and azimuthal temperature gradient (Eq. (8)) and the $\pi/2$ phase shift between the temperature $T$ and $u_r$. Note there is a small offset of ca. 15° between $u_r$ and $\partial T/\partial \phi$, that we attribute to the smallness of the flow Reynolds-number ($Re = Pe = 6.7 \cdot 10^{-4}$). Hence the flow is affected by viscosity and tends slightly to a viscous-buoyancy balance, where $u_r \sim T$. At smaller Ekman numbers this offset is expected to vanish as viscosity is not important in such a limit (Zhang and Gubbins, 1992).

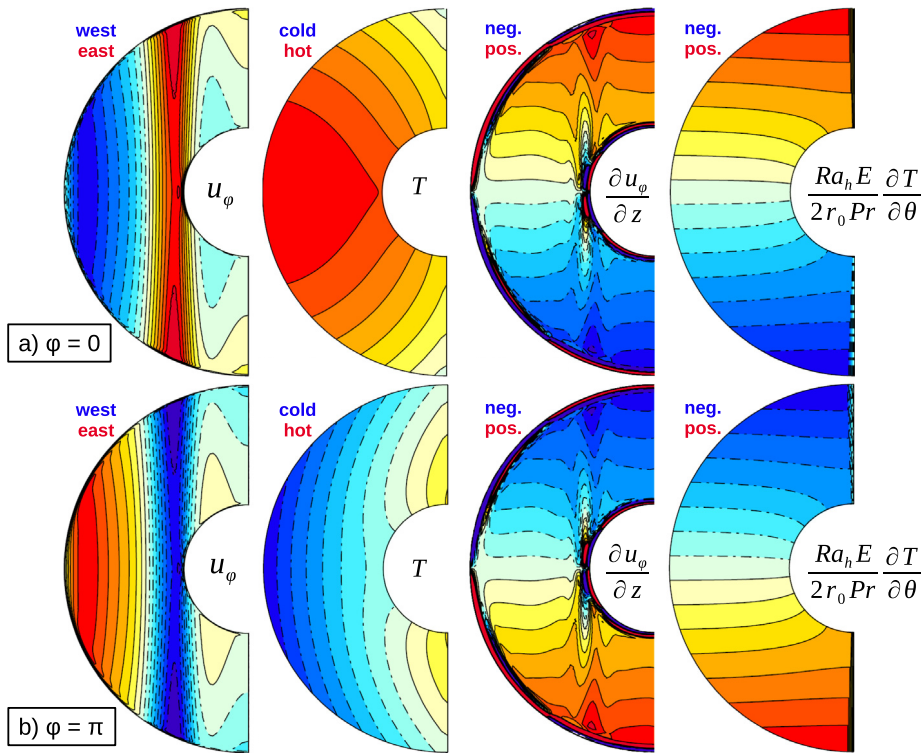

**Fig. 3.** Meridional cuts at $\phi = 0$ (a) and $\phi = \pi$ (b) of azimuthal flow $u_\phi$, temperature $T$ and the two terms of the thermal wind for the conductive case with $Ra_h = 10$. The ratio of Reynolds numbers for geostrophic and ageostrophic flow components are given in Table 1. Parameters: $E = 10^{-4}, q^* = 1.0$.

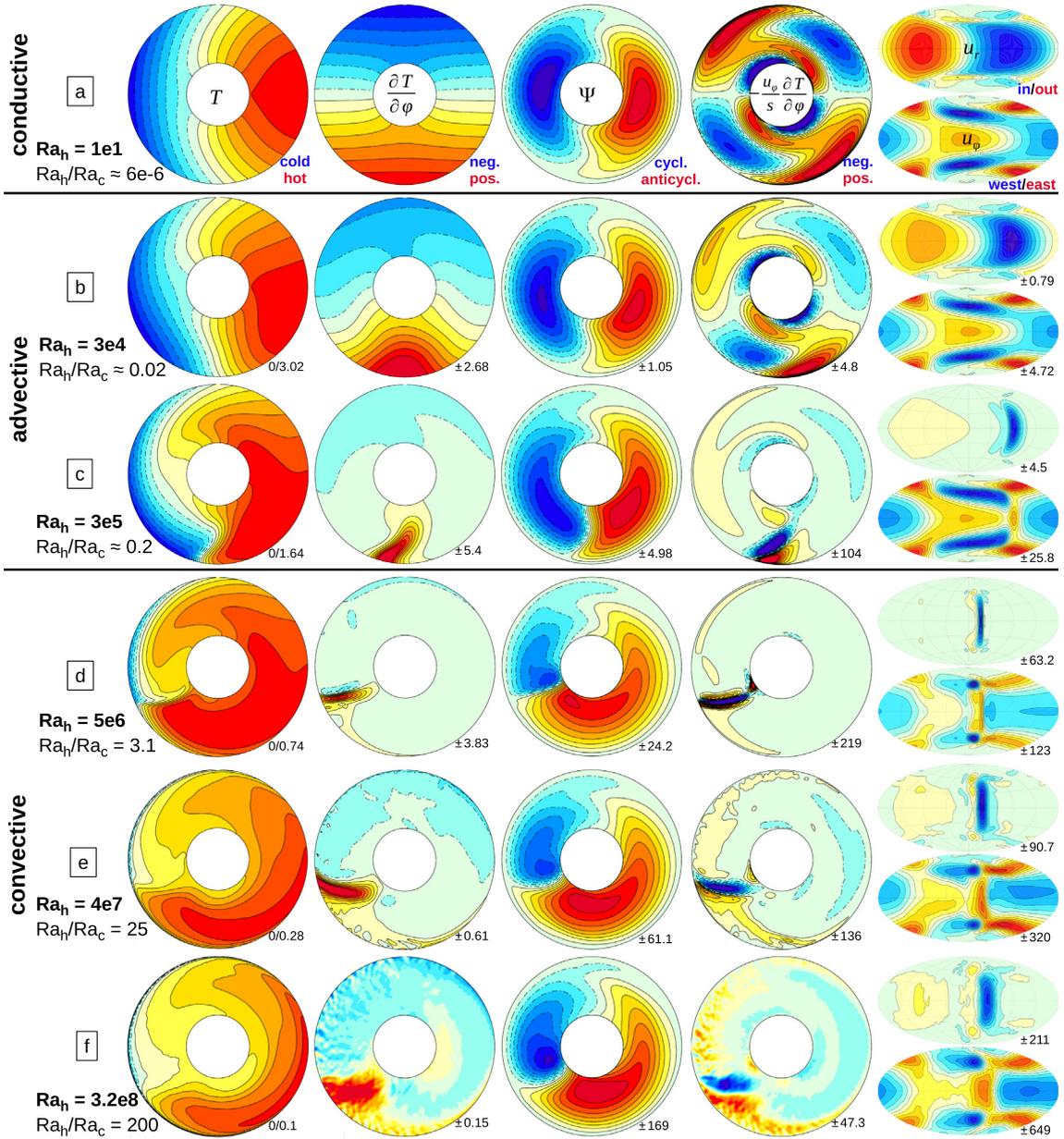

**Fig. 4.** Flow regimes and transitions. Each row shows (from left to right) the time average of temperature, azimuthal gradient of temperature, equatorial stream function, the nonlinear temperature advection term along azimuth in the equatorial plane and radial (upper) and azimuthal flow (lower) on a spherical surface at mid-depth. The (horizontal) Rayleigh number is increased from (a) $Ra_h = 10$, (b) $Ra_h = 3 \cdot 10^4$, (c) $Ra_h = 3 \cdot 10^5$, (d) $Ra_h = 5 \cdot 10^6$, (e) $Ra_h = 4 \cdot 10^7$ to (f) $Ra_h = 3.2 \cdot 10^8$. Contour maxima and minima are indicated at the bottom right in each panel. Note, Table 1 lists ratios of geostrophic and ageostrophic flow amplitudes for the depicted cases. Parameters: $E = 10^{-4}, q^* = 1.0$.

If the driving is increased to $Ra_h = 3 \cdot 10^4$, the symmetry between the eastern and western hemisphere is broken (Fig. 4b). We picked this case, because the mean flow Reynolds-number firstly exceeds unity here ($Re = Pr \approx 1.8$), hence the effect of temperature advection becomes significant. As shown in the plot the downwelling is stronger and more concentrated than the upwelling. This is caused by the azimuthal temperature advection $-u_\phi / s \, \partial T / \partial \phi$, which is added on the right-hand side of the figure. If this term is locally negative (positive) the temperature is increased (decreased). As can be seen from its structure the nonlinear advection provides heating (cooling) in sector I and III at larger (smaller) radii, and vice versa in sector II and IV. Due to the incompressibility, the flows at the larger radii are stronger and global effect is then an increase (decrease) of the azimuthal temperature gradient in the western (eastern) hemisphere. In response, the downwelling (upwelling) in the western (eastern) hemisphere is enhanced (suppressed) and the azimuthal flows are concentrated in the west. At a slightly higher $Ra_h = 3 \cdot 10^5$ (Fig. 4c), the upwelling has almost ceased whereas the downwelling developed a strong and confined jet-like structure, which is inclined in prograde direction while descent.

The azimuthal profiles in Fig. 5 (red and brown) show that temperature flattens out along azimuth, but obeys a strong slope in the vicinity of the jet. This clearly indicates that the jet is separating the cold east from the hot western hemisphere. As discussed before, the local azimuthal temperature gradient is directly responsible and hence indicative for the inward traveling jet. Remarkably, the temperature $T$ looses its phase relation with respect to the radial flows, but the strong anti-correlation between $u_r$ and $\partial T / \partial \phi$ still holds. Hence also for nonlinear regimes, where advection or convection is significant, on time average the boundary induced flow is still fully determined by the temperature structure. Note, this was also successfully tested for the experimental setup of Sumita and Olson (1999), where a model thermal anomaly was applied to find the mean flow structure analytically (Sumita and Yoshida, 2003). In this regime radial temperature (density) variations are enhanced by $Ra_h$, but not sufficiently to reach unstable stratification. As the major nonlinearity stems from the temperature advection this regime might be termed as 'advective'.

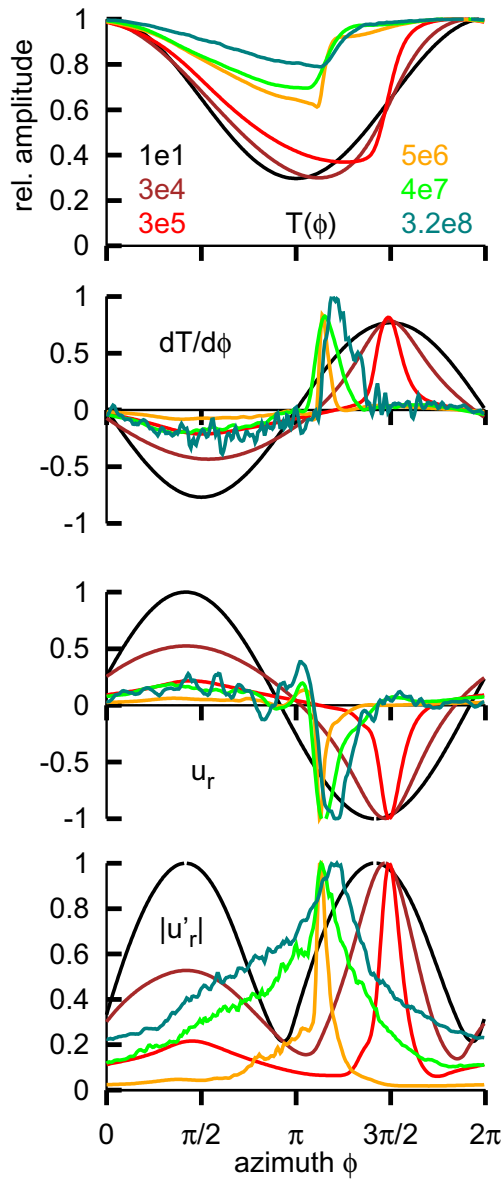

**Fig. 5.** Azimuthal profiles of (from top to bottom) time averaged temperature, azimuthal temperature gradient, radial flow and intensity of radial flow. All curves are normalized to their respective maximal values. Different colour refer to different Rayleigh numbers. Parameters: $E = 10^{-4}$, $q^* = 1.0$. (For interpretation of the references to color in this figure legend, the reader is referred to the web version of this article.)

As the jet sinks in and gains in amplitude with increasing $Ra_h$, the double cyclone structure is deformed and the strong Coriolis force might deflect any inward radial flow into prograde azimuthal direction. Alongside with the jet compression, Fig. 4 shows that a radial dependence of the jet phase shift $\Phi(r)$ is emerging with increasing $Ra_h$ for the advective regime. The phase shift close to the outer and inner boundary, $\Phi_o = \Phi(r_o)$ and $\Phi_i = \Phi(r_i)$, respectively is plotted in Fig. 6, whereas the difference $|\Phi_o - \Phi_i|$ estimates the bending. It can be seen, from $Ra_h = 3 \cdot 10^5$ onwards significant jet bending occurs, as the jet anchor points move prograde in the interior and retrograde close to the outer boundary. The reason for a more retrograde position of the upper boundary anchor point seems to be related to the asymmetry between the cold cyclone and the hot anticyclone. The former features a prograde flow close to the outer boundary, whereas the latter is retrograde. Both flows advect the jet, but the anticyclone is increasingly stronger with higher $Ra_h$ and hence the jet sinks down at smaller $\phi$. On the other hand, close to the inner boundary the azimuthal flow in the dominating anticyclone is prograde yielding a westward shift of $\Phi_i$, which can be seen in Fig. 6, red line. Note, this trend was also observed in the experiments by Sumita and Olson (2002). The total jet bending hence increases with $Ra_h$ until the onset of convection is reached.

Radial convective instabilities driven by unstable buoyancy stratification might participate when the buoyancy scaling factor $Ra_h$ supersedes the critical value $Ra_c$ for the homogeneous model (given in Section 2). Note, it is not clear how the boundary forcing affects the onset of convection. However, we use $Ra_c$ as the regime boundary between advective and convective regime as there are several key properties coinciding. Besides being the critical Rayleigh number in the homogeneous case at $Ra = Ra_c$, we also observe steady flows for $Ra_h < Ra_c$ and fluctuating flows for $Ra_h > Ra_c$ which can be evident when taking snapshots, as shown below in Section 4. The time dependence might be related to both, emergence of time-dependent convective flows and a wiggling of the jet. For strongly convecting models ($Ra_h \gg Ra_c$) at any instance of time, the jet is hardly identifiable but remains persistent after time-averaging. It is thus not the same periodic fluctuations caused by periodic formation and destruction of the front observed in the similar experimental setup by Sumita and Olson (2002).

Furthermore the jet bending $|\Phi_o - \Phi_i|$ is a clear function of $Ra_h$ and has a global maximum at $Ra_h = Ra_c$ (see Fig. 6, vertical line). Note, within the two nonlinear regimes the upper anchor point of the jet moves consistently retrograde to smaller $\Phi_o$ at the outer boundary with increasing $Ra_h$ (Fig. 6). Interestingly, weakly supercritical convection straightens the jet whereas for strongly supercritical cases $\Phi_o$ and $\Phi_i$ seem to settle at 25° and 45°, respec-tively. The jet bending in the advective regime stemmed from the dominant anticyclone yielding a deformation of both cyclonic structures. As the radial jet defines the temperature front, azimuthal bending implies strong radial gradients through the jet. It is reasonable that the radial convective motions setting at $Ra_h = Ra_c$ homogenise radial temperature gradients and only allow for azimuthal gradients to persist. As a consequence the jet is dominantly radial in the convective regime. However for higher $Ra_h \gg Ra_c$ the radial jet is azimuthally deflected by the Coriolis force yielding a weak, but consistent bending.

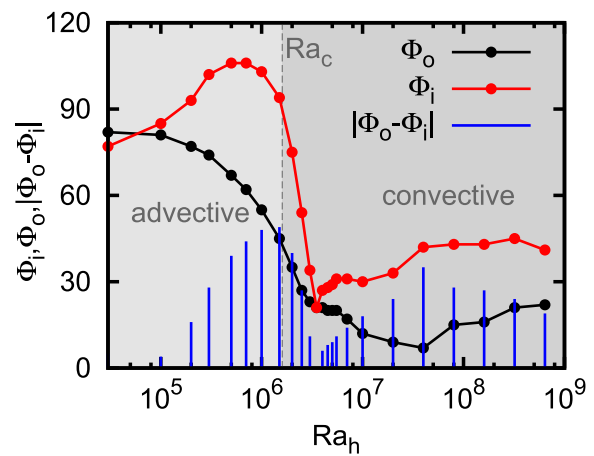

**Fig. 6.** Azimuthal displacement of the inward jet close to the outer and inner boundary, $\Phi_o$, $\Phi_i$, respectively. The vertical blue lines give the difference and hence measure the degree of bending. The grey, dashed vertical line denotes the critical Rayleigh number for the homogeneous case. Parameters: $E = 10^{-4}$, $q^* = 1.0$. (For interpretation of the references to color in this figure legend, the reader is referred to the web version of this article.)

Throughout the convective regime with $Ra_h > Ra_c$, the jet separates the cold cyclone and the hot anti-cyclone and the strongest radial temperature gradients develop (as expected) at the prograde edge of the eastern cyclone. In opposite to the retrograde azimuthal flow in the hot anticyclonic western cell, the prograde azimuthal flow in the eastern cyclone advects hot fluid along a strong azimuthal temperature gradient and hence provides cooling. If buoyancy is allowed and a fluid parcel is sufficiently cooled it can become negatively buoyant due to unstable stratification. Hence, once cooled enough it is the jet itself, that becomes negatively buoyant (heavy) and sinks down at smaller $\Phi$ (further east). Interestingly the buoyancy is not only responsible for the jet to sink in but also cools the eastern cyclone by radial convection. Even for the time-dependent convective regime (Fig. 4e and f), amplitude and position of the jet are perfectly anti-correlated to the azimuthal temperature gradients (Fig. 5, orange and green).

For $Ra_h = 4 \cdot 10^7$ (Fig. 4e) the flow has reached full nonlinearity in the sense, that both radial and azimuthal temperature variations drive flows as the critical Rayleigh number for radial convection is exceeded by far ($Ra_h/Ra_c \approx 25$). The azimuthal profile of the intensity of nonaxisymmetric flows $|u_r'|$ (Fig. 5, bottom plot) shows the dominance of the jet and that the radial convection in the colder, eastern cyclone is much more energetic than in the hotter, western anticyclone. In addition it can be seen, that the stronger the radial convection gets, the closer it is to the dominant jet. As the jet is eastward phase-shifted, the azimuthal profile of $|u_r'|$ does not follow the locally prescribed boundary heat flux (as introduced by the anomaly), but does follow the temperature distribution created by the mean (boundary forced) flow. In other words, the phase-shifting effects of the Sverdrup balance due to rotation also translate into the 'regular' radial convection.

### 3.3. Flow speed scaling

An interesting question is whether the jet amplitudes can be related to the system parameters, such as the boundary forcing amplitude or the rotation rate, and how dominant they are in the presence of additional radial convective instabilities. For the conductive linear regime, that is when the flow is controlled by the boundary anomalies and advective terms are not important, the axial vorticity equation (Eq. (8)) should predict the amplitude of the global circulation. Assuming that the azimuthal temperature gradient is well represented by the boundary forcing Eq. (8) can be approximated by

$$\frac{Re}{\ell} \sim \frac{RaE}{2r_oPr} \frac{\Delta T}{\ell}, \tag{32}$$

where $\ell$ is the variational length scale for flow and temperature. This leads to the flow scaling:

$$Re \sim Ra_hE, \tag{33}$$

where the temperature scale is approximated by the boundary anomaly strength ($\Delta T \approx q^*$).

As a measure of the kinetic energy the global flow kinetic Reynolds number of the time averaged flow is used. Making use of the time-persistence of the boundary induced flows, $Re$ takes radial, azimuthal and longitudinal flows into account but might ignore rapid flow fluctuations such as the convective flow. Fig. 7 tests this scaling for various combinations of $Ra$ (red), $E$ (blue) and $q^*$ (green). The dot-dashed grey line indicates the linear slope and resembles the mean flow amplitude in terms of Reynolds number surprisingly well up to a certain point equivalent to $Ra_h = 3 \cdot 10^4$ (see also Fig. 4, 2nd row). All red cases are calculated with fixed $E = 10^{-4}$, $q^* = 1$ and variable Rayleigh number ranging from 10 to $6.4 \cdot 10^8$. For the blue cases in the figure the Ekman number was halved, and the resulting Reynolds numbers seem to be halved as well. As the third factor of the thermal wind scaling (Eq. (33)), two test cases with halved $q^* = 0.5$ and doubled $Ra$ fall back to mean trend.

It can be seen from the Fig. 7, when reaching $Ra_hE/2 \approx 1$ the linear relation fails and overestimates the flow amplitude. At this point the nonlinear temperature advection term ($\vec{u} \cdot \nabla T$) overcomes the thermal diffusion as indicated by flow the Reynolds or Péclet number (as $Pr = 1$) reaching $Re \approx 1$. We set this as the boundary between linear conductive and nonlinear advective regime. A further increase in Rayleigh number leads to the previously discussed jet focusing and amplification between $Ra_h = 3 \cdot 10^4$ (advective regime) and the onset of radial convection at $Ra_c \approx 1.6 \cdot 10^6$ (convective). For cases with supercritical convection, the flow amplitude seem to be much better resembled by a $Ra_h^{1/2}$-scaling. Such an exponent is theoretically expected for a buoyancy-viscosity balance (King and Buffett, 2013) and well confirmed by numerical models and experiments (Christensen and Aubert, 2006; King et al., 2013). From the axial vorticity equation (Eq. (8)) Sumita and Olson (2002) also suggested a flow scaling $\propto Ra_h^{1/2}$, but this was deduced for the boundary induced azimuthal flow only. We hypothesise, that our global measure is in agreement with their findings as the azimuthal flow dominates the time-averaged kinetic energy. However for the nonlinear advective regime this simple scaling law fails to predict the correct mean flow amplitude.

The kinetic energy corresponding to the two cell flow might be spectrally strongly dominated by kinetic energy contributions obeying a spherical harmonic order of $m = 1$. A measure of the relative strength of ($m = 1$)-flows is given by Hori et al. (2014) as

$$Em1 = \frac{\int_{r_i}^{r_o} \sum_l E_{l,1}^k r^2 dr}{\int_{r_i}^{r_o} \sum_{l,m} E_{l,m}^k r^2 dr}. \tag{34}$$

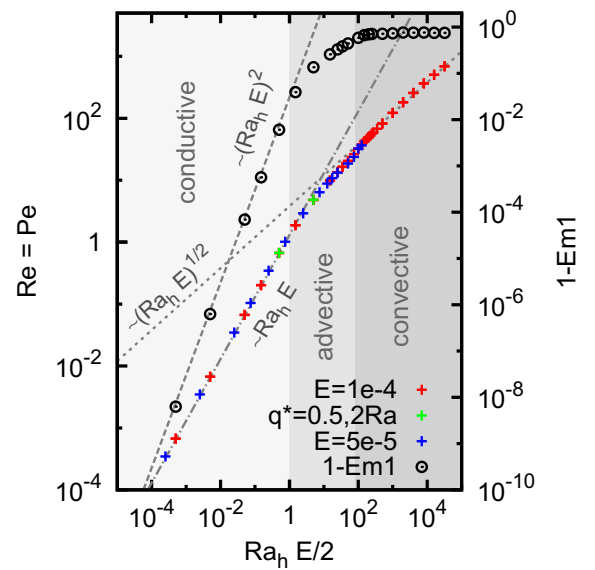

**Fig. 7.** Dependence of the flow amplitude $Re$ (crosses, left ordinate) on Rayleigh and Ekman number and logarithmic deviation from Em1 (circles, right ordinate).

As a consequence of the jet focusing, $Em1$ is expected to decay fast when $Ra_h$ is increased and the spectrum is not as monochromatic (with $m = 1$), but involves higher modes. Fig. 7 shows the deviation from a monochromatic ($m = 1$)-flow. In the plot this quantity increases with a slope of 2 for $Ra_hE/2 < 1$ (conductive regime), undergoes a transition (advective) and saturates to a constant value corresponding to $Em1 \approx 0.24$ for $Ra_hE/2 > 100$ in the convective regime, where radial convection is active. Note, at the first transition between linear conductive and nonlinear advective regime, $Em1$ has only marginally decreased, e.g. at $Ra_h = 3 \cdot 10^4$ we find $Em1 = 0.961$.

## 4. Heat transport in the convective regime

This section contains a comprehensive description of how the radial convection is affected and how the temperature anomaly introduced by the boundary anomaly is extracted from the core. Radial heat transport in a convective flow is given by the radial component of the heat advection term and should usually direct radially outwards. This implies that a positive (outward) radial flow provides cooling if the temperature decreases with radius hence hot material will be advected towards the cooling surface. Note, these flows will be accompanied with the same amount of negative radial flow pumping cool material to the hotter interior (cooling as well). However, if the flow is time-dependent and fluctuating there will be also inward heat flux contributions. We therefore investigate the radial heat transport in a series of individual snapshots of a time-variable flow, collect the contributions of the radial heat transport into outward ($j^o$) and inward ($j^i$) and time-average their intensities. Afterwards the sum of the two might give an idea of the net (outward) radial heat transport.

Locally the direction of the radial heat transport is determined by the sign of the radial temperature gradient, hence

$$\frac{\partial T}{\partial r} \begin{cases} \leqslant 0 : j^o = \langle |u_r \frac{\partial T}{\partial r}| \rangle_t \\ \geqslant 0 : j^i = -\langle |u_r \frac{\partial T}{\partial r}| \rangle_t \end{cases} \tag{35}$$

where both quantities will be present in a fluctuating flow. However, the net radial heat flux

$$j^r = j^o + j^i \tag{36}$$

will be positive as the fluid shell is cooling. The classic Nusselt number as the nondimensional ratio between convective and conductive heat flux is related to the global average of the radial heat transport $\overline{j^r}$ (e.g. Otero et al., 2002) by

$$Nu = 1 + \frac{\overline{j^r}}{\Delta T}. \tag{37}$$

As we use fixed flux boundary conditions, the normalisation temperature scale $\Delta T(Ra)$ is an output quantity. The temperature scale usually drops with increasing convective vigour as stirring and mixing is more efficient.

In accordance to the radial heat transport, we define similarly azimuthal (east/west) and longitudinal (poleward/equatorial) heat transport by

$$j^a = j^e + j^w \tag{38}$$

$$j^l = j^p + j^{eq}. \tag{39}$$

Presumably for a model with homogeneous outer boundary heat flux there will be no net azimuthal or longitudinal heat transport. Although the individual directions will be significant, on time-average they will cancel each other out.

### 4.1. Convective structures

We plot snapshots of radial ($u^r$) and azimuthal flow ($u^\phi$) and heat transport along radius ($j^r$) and azimuth ($j^a$) in the equatorial plane for several combinations of convective driving $Ra$ and forcing amplitudes $q^*$ in Fig. 8. For the homogeneous case the convective instability sets in at $Ra_c = 1.6 \cdot 10^6$ with an azimuthal wavenumber of $m_c = 13$ (Fig. 8, panel a). This is in line with the results of Hori et al. (2012), where a similar system was studied. Note, at $Ra = Ra_c$, the system is steady and hence there is only a positive contribution to the radial heat flux ($j^r$). The convection sets in at mid-depth. Note for this special case, the very weak azimuthal contribution $j^a$ was enhanced by a factor of 50 relative to the radial transport to visualise the structure better. For all other cases in Fig. 8, in each row $j^r$ and $j^a$ share the same contour levels.

Applying a boundary forcing with $q^* = 1.0$ changes the flow drastically, where an azimuthal wavenumber of $m = 1$ is dominant. The radial and azimuthal flows (Fig. 8b) resemble the time average structures discussed before (Section 3.2). The compressed inward jet harbours already both, inward and outward radial flow. The azimuthal flows feeding into the jet start to contribute to the redistribution of heat. Especially the hot anticyclonic flow transports a large amount of heat towards the jet. For this case, the boundary induced flows are much stronger than the weak radial convection.

Increasing the convective supercriticality slightly to $Ra = 5 \cdot 10^6$ ($Ra/Ra_c = 3.13$, Fig. 8c; cf. Fig. 4d) and keeping $q^* = 1.0$ gives rise to radial convective flows significantly participating in the flow. Especially stronger convection and hence more efficient cooling develops west of the jet, but is clearly suppressed in the opposing hemisphere. The radial outward heat transport supported by the convective flows is now strongest close to the outer boundary. The azimuthal transport is equally partitioned between the eastward advection (positive, red colours) west of the jet and westward advection (blue colours) east of the jet. That implies both azimuthal flows advect heat towards the jet.

At an even higher $Ra = 4 \cdot 10^7$ ($Ra/Ra_c = 25$, Fig. 8d; cf. Fig. 4e) and $q^* = 1.0$ the flow has developed full time-dependence. Radial convection fills most of the left half (sectors II and III) of the shell, but seems most energetic west of the jet where the temperatures are colder and hence stronger convection can occur. The radial heat flux is outwards everywhere, but the sinking jet clearly deposits heat into the interior as well (blue) and hence heats up the areas beneath the anticyclone, which suppresses convection. The azimuthal direction is more mixed between eastward and westward, but is still mainly eastward (westward) in front of (behind) the jet.

In Fig. 8(e), we reduced the forcing amplitude to $q^* = 0.5$, where the jet is still clearly identifiable, though its position and most likely its power have been altered. It is quite remarkable, that the convective vigour and hence the net radial heat transport remain a strong function of azimuth, even though the flow is clearly supercritical everywhere.

As a reference case, we also add $Ra/Ra_c = 25$ with no anomalous heat flux in Fig. 8(f). Here the entire shell is filled homogeneously with turbulent convection. There is no net azimuthal transport, whereas the radial one shows the expected behaviour of dominantly outward radial heat transport (net cooling) with weak azimuthal variation.

### 4.2. Radial, azimuthal and longitudinal heat transport

To show the concept of the three components of $j$ we compare two almost identical cases characterised by fixed parameters ($Ra = 4 \cdot 10^7$), but one with homogeneous ($q^* = 0$) outer boundary heat flux and one with boundary forcing ($q^* = 1.0$). The convective supercriticality for this setup is $Ra/Ra_c = 25$ hence we expect rich nonlinear dynamics with strong time variability. Correlating the heat transport over a set of 30 individual snapshots might provide a confident mean value. Fig. 9 plots the radial, azimuthal and longitudinal (upper, middle, bottom) heat transport averaged over $r$ and $\theta$ as function of azimuthal angle.

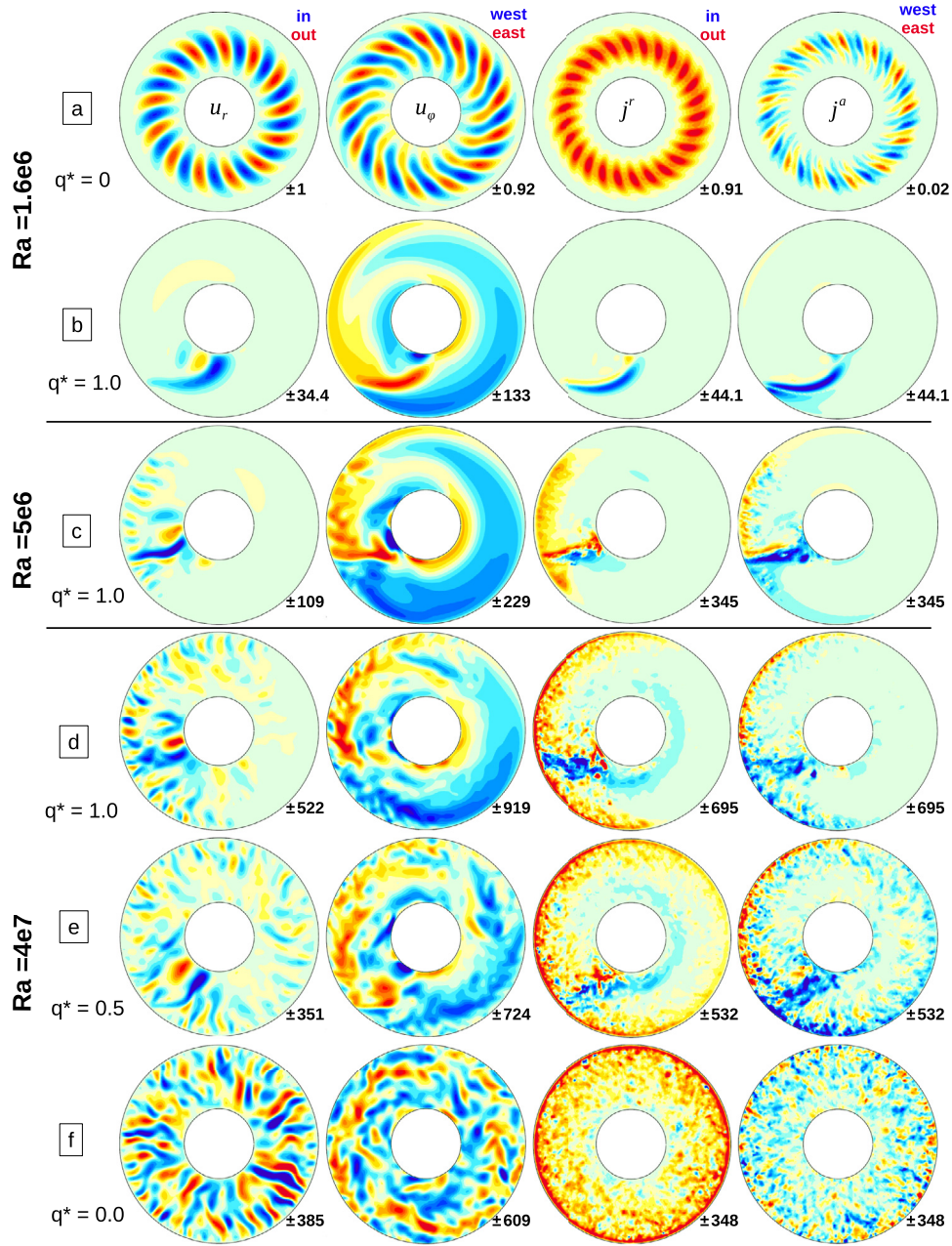

**Fig. 8.** Convection and heat transport properties. Each row shows (from left to right) a snapshot of radial and azimuthal flow ($u_r$, $u_\phi$) and the net radial and azimuthal heat transport ($j^r$, $j^a$) in the equatorial plane. The colours (red–blue) refer to outward–inward ($u_r$, $j^r$), eastward–westward ($u_\phi$, $j^a$). The reference Rayleigh number and the boundary forcing amplitude are listed. Note, the azimuthal heat transport for $Ra/Ra_c = 1.0$ and $q^* = 0.0$ (top plot) is marginal in that case and amplified by a factor of 50 for visualisation purposes. Parameters: $E = 10^{-4}$, the convective supercriticalities are $Ra/Ra_c = 1$ in (a and b), $Ra/Ra_c = 3.125$ in (c) and $Ra/Ra_c = 25$ in (d–f). (For interpretation of the references to colour in this figure legend, the reader is referred to the web version of this article.)

For the homogeneous case (light blue, blue and dark-blue) the outward and inward heat transport profiles are rather flat and sum up to a net outward radial transport. This reflects the action of a fluctuating convective flow. There is no net horizontal heat flux. However, if the variable outer boundary heat flux is applied, a strong azimuthal dependence and net horizontal heat transport develops. As a reminder, the minimum (maximum) heat flux at the outer boundary is at $\phi = 0$ ($\phi = \pi$). The large peaks in all curves for $q^* = 1.0$ (orange, red and dak-red) align well with the phase shifted inward jet and are not anchored where the heat flux is maximal but are azimuthally displaced by ca. 30° further east. Also, it can be clearly seen, that each of the profiles is strongly suppressed where the heat flux is smaller. The net radial heat transport features a maximum slightly westward of the jet and minimum slightly eastward of the jet. This represents nicely the enhanced cooling efficiency in the colder eastern cyclone and thermal blanketing in the hot western hemisphere. The jet diverts the azimuthal flows downwards and hence transports both hot and cold fluid to the interior (see also Fig. 8d). Note, on an azimuthal average the amount of heat transported outwards is less in the boundary forced than in the homogeneous case and hence the core shell remains hotter.

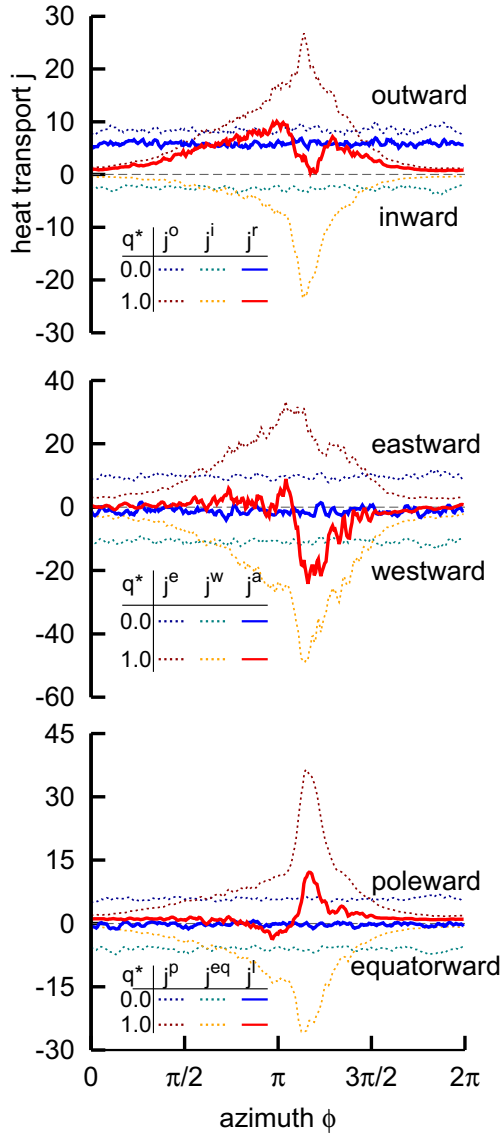

**Fig. 9.** Heat transport for $Ra = 4 \cdot 10^7$ along radius, azimuth and longitude (top, middle, bottom panel). Homogeneous case ($q^* = 0.0$, bluish), boundary forced one ($q^* = 1.0$, reddish). For each panel the individual contributions (e.g. inward/outward) are thin dashed lines, where the net heat flux is plotted with thick solid lines. The indices relate to heat transport o/i/r – outward/inward/net radial, e/w/a – eastward/westward/net azimuthal, p/eq/l – poleward/equatorward/net longitudinal. (For interpretation of the references to colour in this figure legend, the reader is referred to the web version of this article.)

The boundary forcing stimulates strong azimuthal flows towards the jet, hence we expect a significant contribution of azimuthal heat transport when $q^* > 0$. The middle plot of Fig. 9 proves that both, the cyclone and the anticyclone, participate into a net azimuthal flux. It can be seen, that even in the net azimuthal heat transport ($j^a$), the transport is (weakly) eastwards further west of the jet (smaller $\phi$) and westward further east of the jet. Both converge into the jet. However, as the eastern cyclone is more isothermal due to the enhanced radial convective flows the anticyclonic westward transport is stronger and hence the net azimuthal flow is westward and peaks at the eastern edge of the jet.

Interestingly also a net longitudinal heat transport is created by the boundary anomaly. This transport is aligned with the jet and directed polewards. As at any radial level the ageostrophic part (Eqs. (23) and (24)) dictate the strong radial inward flow of the jet to coincide with strong polarward flow, this suggests a decreasing temperature towards the poles which is found in our models (not shown).

### 4.3. Nusselt number vs. Rayleigh number

We expect that when convection becomes more vigorous at higher Rayleigh number the efficiency of stirring and mixing is strongly enhanced, and hence the influence of any thermal boundary inhomogeneity might diminish. We therefore calculate the net radial heat transport in terms of the Nusselt number (Eq. (37)) for cases between weakly nonlinear with $Ra = 10^7$ up to strongly turbulent $Ra = 6.4 \cdot 10^8$, ranging in terms of convective supercriticality from $6.25 \leqslant Ra/Ra_c \leqslant 400$. Averages are taken over radius and longitude hence the $j^r(\phi)$-curves are similar to the top plot in Fig. 9. The Rayleigh numbers used for Fig. 10 are stepwise increased by factor four and $q^* = 1$ is fixed. Note, for visualisation purposes the thick, bright curves are smoothed using gnuplot's standard Bezier smoothing function, whereas the darker and thinner curves give the real data. Interestingly, the enhanced radial convection between $\phi = [\pi/2 \ldots 3\pi/2]$ (or in sectors II and III) and the suppression on the opposite half seems quite independent of the Rayleigh number. Also for all cases the convection is stronger (weaker) west of the jet, and weaker further east. The independence on the convective vigour $Ra_h$ is a quite remarkable result, as for rather high $Ra/Ra_c$ the flow is almost supercritical everywhere.

The suppression of the net radial heat transport in sectors I and IV should be visible as well in the Nusselt numbers. We calculate $Nu$ with Eq. (37) for a set of 13 different Rayleigh numbers for the homogeneous reference cases $Nu^0$ (Fig. 11, red), for the equivalent boundary forced models $Nu^1$ (black) and their ratio (blue). The smallest is the critical Rayleigh number for the homogeneous reference case. By definition $Nu^0$ should be marginal there (black curve), but the boundary forcing actually enhances the heat transport (red). It is expected by linear theory that at the onset any boundary anomaly will enhance $Nu$ (Kelly and Pal, 1978). However, increasing the Rayleigh number towards more nonlinear systems shows that at any supercritical test case the homogeneous case ($Nu^0$) features a higher Nusselt number or more efficient heat transport than the boundary forced case ($Nu^1$). The suppression by the thermal boundary anomaly seem to settle at ca. 50% for strongly driven models. Given the sparse data, we do not attempt to derive a scaling relation like $Nu \sim Ra^\alpha$. However, it is well accepted that $\alpha$ is smaller than unity for weakly rotating systems and depends in the model setup etc. E.g., King et al. (2010) suggest $\alpha = 2/7$ for a weakly rotating spherical shell with fixed temperature contrast and a dynamo process, whereas a comparable model but with rapid rotation follows $\alpha = 6/5$. We find the rapid rotating behaviour only for models very close to the onset of convection. For the majority of our models $\alpha \approx 0.45$ seems reasonable, when $Ra_h \gg Ra_c$ as indicated in Fig. 11.

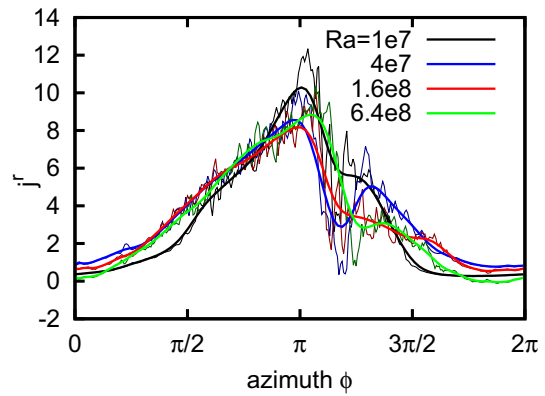

**Fig. 10.** Azimuthal variation of net radial heat transport $j^r(\phi)$ for various Rayleigh numbers. Note, for better visibility the thick bright curves are smoothed, whereas the darker thin curves give the numerical results.

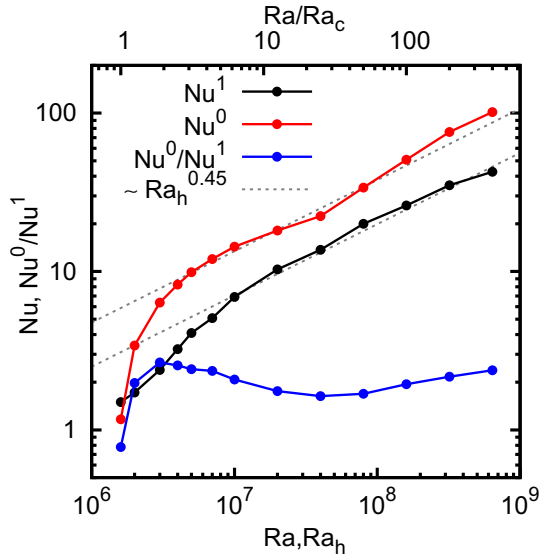

**Fig. 11.** Nusselt number for $q^* = 1.0$ (black) and $q^* = 0.0$ (red) as function of Rayleigh number. The ratio is given in blue. (For interpretation of the references to colour in this figure legend, the reader is referred to the web version of this article.)

The reason for the suppression of convective efficiency by the boundary forcing is then directly related to that exponent. As a toy model of the boundary forced cases, we assume the Rayleigh number, which is determined by the radial temperature gradient, shall be doubled in one half of the core and set to zero in the other one. This leads to different values of the globally-averaged radial heat transport, i.e. the Nusselt number. Compared to a model with homogeneous boundary heat flux, the heat transport can be given as

$$\frac{Nu(Ra)}{\frac{1}{2}Nu(2Ra)} = \frac{Ra^\alpha}{\frac{1}{2}2^\alpha Ra^\alpha} = 2^{1-\alpha}, \quad (40)$$

yielding a reduction of heat transport in the boundary forced case for $0 < \alpha < 1$. Note, the suppression measured in the numerical models is much stronger than this simplistic estimate ($\approx 1.46$ for $\alpha = 0.45$) as the ratio $Nu^0/Nu^1$ seems to be bordered by 1.8 and 2.5. Even though areas with reduced local $q$ are exactly compensated with enhanced $q$ areas and hence on zonal average the total heat flux is identical between a ($q^* = 0.0$)- and a ($q^* = 1.0$)-model, the boundary forcing reduces the net radial heat transport independently of the convective vigour. Physically this implies, that the additional azimuthal heat transport introduced by the boundary anomaly consumes a significant fraction of the available convective kinetic energy, hence the mean radial heat transport is reduced.

### 4.4. Dependence on heat flux anomaly amplitude

As a final set of models, we keep all parameters fixed ($Ra = 4 \cdot 10^7, Ra/Ra_c = 25, E = 10^{-4}$) and vary $q^*$. Such a procedure will test the sensitivity of a vigorously convecting system regarding inhomogeneities at the CMB heat flux. Starting with the azimuthal profiles of the net radial heat flux $j^r$ in Fig. 12, we conduct numerical experiments for a few other $q^*$-values. Note, only the model related to $q^* = 1.0$ (black curve) features a neutrally critical heat flux at $\phi = 0$. However, the smooth transition between $q^* = 0$ (grey) and $q^* = 1.0$ (black) suggests a rather linear dependence on $q^*$. Note, that the position of the jet indicated by the rough decay of $j^r$ is shifting towards larger $\Phi$ and the amplitude is weaker with decreasing $q^*$.

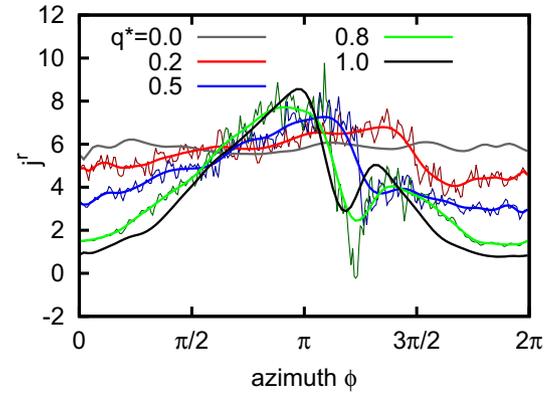

**Fig. 12.** Azimuthal variation of net radial heat transport $j^r(\phi)$ for fixed $Ra = 4 \cdot 10^7$ and various $q^*$ values. Note, for better visibility the thick bright curves are smoothed, whereas the darker thin curves give the numerical results.

For measuring the azimuthal asymmetry of the radial heat transport we define a hemisphericity like

$$H_j = \frac{|j^r_{max} - j^r_{min}|}{|j^r_{max} + j^r_{min}|}, \quad (41)$$

where the maximal $j^r_{max} = \max[j^r]$ is taken between $\pi/2 \leqslant \phi \leqslant 3\pi/2$ (sectors II and III) and the minimal $j^r_{min} = \min[j^r]$ from the other half. Fig. 13 collects the results of $H_j$ and the Nusselt numbers according to Eq. (37). It can be seen, that both quantities depend almost linearly on $q^*$.

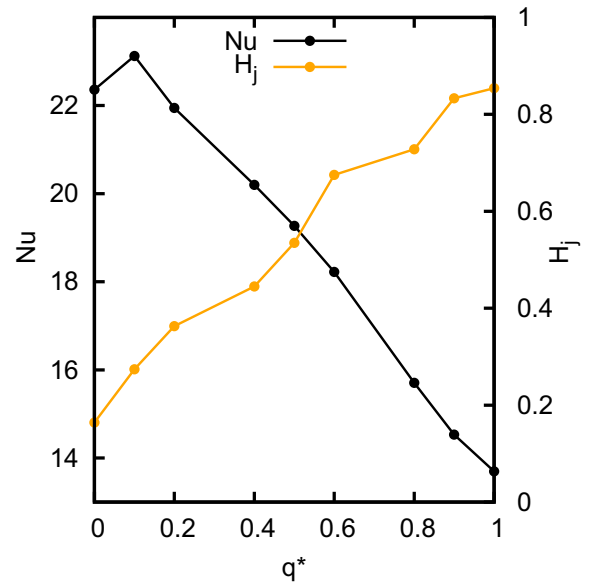

**Fig. 13.** Nusselt number for $Ra = 4 \cdot 10^7$ and the azimuthal asymmetry of the radial heat transport $H_j$ (orange) as a function of boundary anomaly amplitude $q^*$ (black). (For interpretation of the references to colour in this figure legend, the reader is referred to the web version of this article.)

Hence we suggest that the effects on the convection do not depend at all on the convective vigour and only linearly on the boundary forcing amplitude.

## 5. Conclusions

We numerically investigated core flows under inhomogeneous cooling at the outer boundary. When the core is subcritical to radial thermal convection, these thermal anomalies induce core flows whose linear and nonlinear properties are discussed in detail. We also study the situation when, in contrast, the core is unstable to thermal convection and a non-homogeneous CMB heat flux will strongly alter the convective structures and the heat transport. To model the horizontal variation of the lower mantle temperature in a terrestrial planet, we apply a perturbation of the CMB heat flux shaped as a spherical harmonic $Y_{11}$ and perform a comprehensive study with a numerical model of an incompressible and viscous fluid heated from within and cooled from above, which is contained in a rapidly rotating spherical shell. Such a simple heating setup takes only secular cooling into account and hence represents best the liquid cores of the Early Earth, Mars or any terrestrial exoplanet before inner core solidification. As the main control parameters we conduct a parameter survey regarding the Rayleigh number $Ra$ and forcing amplitude $q^*$. In the absence of radial convection ($Ra_h = Raq^* < Ra_c$), a linear *conductive* regime (determined by the vorticity balance between heat conduction and Coriolis effects and the ageostrophic thermal wind) and a nonlinear *advective* regime controlled by thermal advection are identified. If $Ra_h > Ra_c$ radial convection participates strongly in the nonlinear dynamics of a *convective* regime.

In general such a heat flux anomaly introduces a global azimuthal and latitudinal temperature gradient, which tends to be equilibrated by a dominantly geostrophic mean two-cell circulation given by a broad cyclonic structure in the eastern hemisphere and an anticyclone in the west, which converge into a radial inward flow. As it was previously suggested and in agreement with the linear theory, see Section 3.1 in this paper (or Zhang and Gubbins, 1992; Yoshida and Hamano, 1993) the radial flows are not located where the outer boundary heat is maximal, but where the azimuthal temperature gradient is maximised. For the linear model this phase shift is 90° eastward or a quarter of the azimuthal wavelength of the anomaly. This can be understood as the Sverdrup relation, which is well-known in the context of geophysical fluid dynamics.

Starting from a non-convecting, and almost linear numerical model ($Ra_h = 10$), the numerical findings confirm the analytical predictions of the induced geostrophic and ageostrophic flows. For higher $Ra$ and once the advective terms become dynamically important, the radial inflow is compressed into a jet, bended in prograde direction and moves retrograde to smaller phase shift angles. The nonlinearities in the system heat up the western and cool the eastern side of the core. It might be stated, in the absence of convection any inhomogeneously cooled system drives baroclinic flows, whose nonlinear properties, such as the effect of advection or heat transport were analysed here.

If the driving is enhanced beyond the critical value for the onset of convection in the equivalent homogeneous model ($Ra_h > Ra_c$), radial convective flows appear and are enhanced where the heat flux is increased, and strongly suppressed where the heat flux is weaker than average. This convective asymmetry is a consistent property, emerges independent of the convective vigour and increases linearly with the heat flux anomaly amplitude $q^*$. It could be also shown, that the convection is most energetic west of the inward jet where cooling is most efficient.

We also investigated the properties of heat transport, and find that the boundary anomaly introduces a net azimuthal heat trans-port towards the descending jet. As a consequence of the laterally variable cooling efficiency, the Nusselt number representing the global radial heat transport is always smaller for boundary forced than for homogeneously cooled models. Implying that the mean radial convective heat transport is globally reduced by at least 50%, although the azimuthal averaged heat flux is the same in both models. The amount of kinetic energy available for radial convection is reduced by the boundary driven azimuthal flows and the inward jet. As a consequence, the liquid iron cores of terrestrial planets facing inhomogeneous cooling might be hotter and much less well mixed. In addition core convection and the potential induction of global magnetic field might last longer. Remarkably, the local suppression of core convection is rather independent of convective vigour (Rayleigh number), but is linearly amplified by the relative strength of the boundary CMB inhomogeneity.

It seems contradictory, that in a series of comparable studies (Sreenivasan, 2009; Aurnou and Aubert, 2011; Dietrich and Wicht, 2013), the magnetic field induction process was amplified or triggered by CMB heat flux anomalies, whereas our results suggest a strong suppression of core convection for any anomalous heat flux. It has been shown in our models that the boundary forcing drives additional flows favouring a supercritical dynamo, where it would fail for a homogeneous system. The local suppression and amplification of convective flows in response to the weaker or stronger CMB heat flux will only lead to an azimuthal variation of the magnetic field intensity, but not terminate the dynamo. Indeed the local enhancement makes a dynamo more likely. It might be also said, that equatorially antisymmetric and axisymmetric heat flux anomalies drive fierce axisymmetric flows which provide shearing hence strengthen the dynamo process.

Our models are applicable to terrestrial exoplanets orbiting their host star in synchronous rotation. Assuming a similar front-to-back thermal anomaly at the CMB as imprinted on the surface by the irradiation of the host star, the Sverdrup balance yields significant azimuthal flows on the far side converging into a prograde spiralling inward jet. If the liquid core is further convecting, a potential magnetic field induction process might be strongly concentrated on the far side of the planet.

Compared to experimental works by Sumita and Olson (1999) our models reproduce several key features, such as the development of a sharp temperature front located east of the heat flux anomaly and hosting the radial inward spiralling jet and separating the cold east from the hot west. Furthermore the experiments reported a scaling relation suggesting the amplitude boundary induced flows increases like the square root of the total anomalous heat flux through the anomaly (Sumita and Olson, 2002), which is confirmed within our numerical results. This implies the very efficient turbulent stirring and mixing due to the regular radial convection in the core might not suppress the influence of boundary anomalies.

## Acknowledgements

The authors like to thank the referees, Hagay Amit and Ikuro Sumita for constructive comments significantly improving the manuscript. Further we acknowledge Robert Teed providing the routine to transform from spherical to cylindrical coordinates. This work was undertaken on ARC1, part of the High Performance Computing facilities at the University of Leeds, UK. WD is supported in part by the Science and Technology Facilities Council (STFC), 'A Consolidated Grant in Astrophysical Fluids' (reference ST/K000853/1). KH is supported by the Japan Society for the Promotion of Science (JSPS) under a grant-in-aid for young scientists (B) No. 26800232 and by JSPS Postdoctoral Fellowships for Research Abroad.